\documentclass[12pt]{article}

\usepackage{a4wide}
\usepackage{amsmath}
\usepackage{amsfonts}
\usepackage{amssymb}
\usepackage{cite}

\makeatletter

\@addtoreset{equation}{section}
\makeatother

\newcommand{\cd}{\!\cdot\!}
\def\half{\frac12}
\def\d{\mathrm{d}}

\def \be { \begin{equation} }
\def \ee { \end{equation} }
\def \beal#1 { \begin{align} #1 \end{align} }
\def \nn {\notag\\}

\begin{document}

\begin{titlepage}

\title{
\vspace{-2cm}
       \vspace{1.5cm}
       Self-gravitating equilibrium with slow steady flow and its consistent form of entropy current
       \vspace{1.cm}
}
\author{
Shuichi Yokoyama\thanks{syr18046\textcircled{a}fc.ritsumei.ac.jp},\; 
\\[25pt] 
{\normalsize\it Department of Physical Sciences, College of Science and Engineering,} \\
{\normalsize\it Ritsumeikan University, Shiga 525-0058, Japan}
}

\date{}

\maketitle

\thispagestyle{empty}

\begin{abstract}
\vspace{0.3cm}
\normalsize

A relativistic self-gravitating equilibrium system with spherical symmetry as well as with steady energy flow is investigated perturbatively around the hydrostatic limit, where the radial component of the fluid velocity field $u^\mu$ is sufficiently small. 
Each component of vectors and tensors consisting of the system is expanded in different powers, which makes the covariant perturbation approach ineffective. 
The differential equations to determine the subleading correction of the structure variables are presented.
The system retains the current $j^\mu$ accounting for the steady flow, which contributes to the entropy current $s^\mu$ in such a general covariant form that $s^\mu=au^\mu+ bj^\mu$ with $a, b$ unknown parametric functions. 
To determine them, a new condition is proposed.
This condition imposes the entropy current to be of an unconventional form $s^\mu=(s-bj^0)u^\mu/u^0+ bj^\mu$, where $s$ is the entropy density. 
The remaining parameter $b$ is fixed by the current conservation equation. The perturbative analysis shows that $b$ starts with the quadratic order and its leading term is determined explicitly.

\end{abstract}
\end{titlepage}

\section{Introduction} 
\label{Introduction} 

As the constellation has been known from very early days, the same stars have been shining for a very long time. 
In order to address a mystery of such stable radiant emission of astronomical objects, physicists have studied self-gravitating equilibrium systems modeling them, particularly the ones consisting of fluid due to their origin coming about from gas and dust floating in our universe.
Due to the development of nuclear physics in the early twenty century, 
the origin of the long-lasting radiant emission of a star was discovered to be a series of subatomic chain reactions occurring deep inside
\cite{eddingtoninternal,PhysRev.53.595,Bethe:1939bt}. 
Such internal structure of astronomical objects has been studied extensively by employing the Newton's universal law of gravity until the present. (See textbooks \cite{harwit1988astrophysical,phillips1994physics,tayler1994stars,rnaasaaaa67bib2,Kippenhahn2012,choudhuri2010astrophysics}.)
However, since huge energy production due to nuclear chain reactions is nothing but the effect of the Einstein's theory of relativity \cite{1905AnP...323..639E}, the crucial discovery about stellar interior implies that the theory of general relativity is more suitable to explore deep interior of luminous stars even with medium density such as main sequence ones \cite{Yokoyama:2023gxf}. (See also \cite{Yokoyama:2024wad}.)

The general relativistic study of a hydrostatic equilibrium system with spherical symmetry was performed to explore physics of compact stars and local thermodynamics \cite{PhysRev.35.904,PhysRev.36.1791,PhysRev.55.374}. (See textbooks \cite{1934rtc..book.....T,alma991013077249704706,hawking1973large,Misner:1974qy,hartle2003gravity}.) 
What has been uncovered is the relativistic effect in the detailed balance equation for the static equilibrium known as the Tolman-Oppenheimer-Volkov (TOV) equation and the equivalent relation between the gravitational potential and the local temperature known as the Tolman relation. Indeed it was recently shown in \cite{Yokoyama:2023nld} that the Tolman temperature arises both in the Euler's relation and the first law of thermodynamics holding locally and non-perturbatively. (See also \cite{Oppenheim_2003}.) This was done by refining the method to construct the entropy current as a non-Noether conserved current proposed in \cite{Aoki:2020nzm}. (See also \cite{Yokoyama:2023kkg,Aoki:2023ufz,Diakonov:2025wtt}.) 

What was found in the recent research is that the conventional relation of entropy current and entropy density assumed in the analysis of relativistic fluid system as also seen in reviews and textbooks, for example, \cite{landau1987fluid,Gourgoulhon:2006bn,Andersson_2007,rezzolla2013relativistic}, is not right in general in light of the transformation rule of the total entropy \cite{planckRT} as well as of the consequence leading to its non-conservation in such a relativistic equilibrium system.
This result indicates any consequence drawn by using the conventional relation of entropy current and entropy density to be reexamined or corrected, while it is natural to ask how the correct form of entropy current is obtained if it exists and plays any physical role.
This question is important to be answered with taken into account a pioneering work to develop a general framework of relativistic fluid dynamics in curved spacetime by postulating the existence of entropy current \cite{Israel:1979wp}. 
This issue was addressed in exploring a microscopic foundation of entropy current and its quantum origin \cite{PhysRevD.99.125011} (see also \cite{BECATTINI2024138533}) by assuming a certain relation of the partition function and the density operator \cite{Zubarev:1979afm}, while there was research in a standpoint for fluid systems to be analyzable without relying on entropy current \cite{Banerjee_2012,Jensen_2012}. 

The motivation of this research is to address this question by exploring a new self-gravitating equilibrium fluid system discovered recently by the author \cite{Yokoyama:2025gbi}. 
The system is obtained by extending the classic hydrostatic equilibrium system so as to include steady energy flow modeling the stable radiant emission of a star without breaking spherical symmetry. 
The steady energy flow is accounted for by an additional current to the steady fluid system, and this current is expected to contribute to the entropy current in the system. 
A main goal of this paper is to clarify how to determine the contribution of the additional current to the entropy current and to derive its expression explicitly. 
To the end, starting with a review of the previous study of a self-gravitating equilibrium system with steady flow, the perturbative analysis of the system is presented in the following section. 
Subsequently we move on to the analysis of the entropy current and end with discussion.

\section{Self-gravitating equilibrium with steady flow}

\subsection{Review of exact results} 
\label{ExactResult}

The system to be investigated is the relativistic self-gravitating equilibrium one with spherical symmetry and steady energy flow. This was constructed by turning on a current accounting for steady energy flow to the classic hydrostatic equilibrium system \cite{Yokoyama:2025gbi}. Main results are reviewed below.   

The classic relativistic hydrostatic equilibrium system is reviewed below \cite{PhysRev.35.904,PhysRev.36.1791,PhysRev.55.374}. (See textbooks \cite{1934rtc..book.....T,alma991013077249704706,hawking1973large,Misner:1974qy,hartle2003gravity}.)
The line element is given by 
\beal{
\mathring g_{\mu\nu} \d x^\mu \d x^\nu
= -\mathring f (\d x^0)^2 +\mathring h \d r^2 +r^2 \tilde g_{ij} \d x^i \d x^j, 
\label{LE0}
}
where $\tilde g_{ij}$ is the metric tensor for a $d-2$ dimensional Einstein manifold and $\mathring f,\mathring h$ are functions only of the radial coordinate $r$, while 
the energy-momentum tensor is a perfect fluid 
\beal{
\mathring T^{\mu\nu} = (\mathring p + \mathring \rho)\mathring u^\mu \mathring u^\nu +\mathring p \mathring g^{\mu\nu} ,
\label{PerfectFluid}
}
where $\mathring u^\mu$ is the normalized fluid velocity with $\mathring g_{\mu\nu} \mathring u^\mu \mathring u^\nu=-1$ and $\mathring p, \mathring \rho$ are scalar functions only of $r$. 
The Einstein field equation encodes the detailed balance equation referred to as the TOV one \cite{PhysRev.55.374} and the laws of relativistic local thermodynamics and fluid dynamics as well. 
The local temperature $\mathring T$ was discovered to be given by by the Tolman relation as $\mathring T=T_\circ e^{-\mathring\phi}$, where $T_\circ$ is an integration constant and $\mathring\phi$ is the gravitational potential $\mathring\phi =\half\log\mathring f$ \cite{PhysRev.35.904,PhysRev.36.1791}. The integration constant $T_\circ$ needs to be positive due to the positivity of temperature. 
Indeed, this local temperature was shown to appear in both the local Euler's relation and the first law of thermodynamics, in which the entropy density was determined as a charge density read from the non-Noether conserved current \cite{Yokoyama:2023nld}. The construction of the conserved current is explained  in section~\ref{EC}.

In order to extend the results in the above hydrostatic equilibrium system by adding steady flow, it is necessary to include the contribution of a current $j^\mu$ accounting for steady flow to the energy-momentum tensor of the perfect fluid. The minimal inclusion of the contribution in a way to respect the general covariance leads to 
\beal{
T^{\mu\nu} = (\check p + \check \rho)u^\mu u^\nu +\check p g^{\mu\nu} + u^{\nu} q^{\mu}+ u^{\mu} q^{\nu}, 
\label{EMT}
}
where $q^\mu$ is related to the current $j^\mu$ as $q^\mu=\chi j^\mu$ with the prefactor $\chi$ a scalar function of $r$.%
\footnote{ 
In \cite{Yokoyama:2025gbi}, $\chi$ was fixed as $\chi=T$ by using an effective equation of state in the hydrostatic case. This result is not exact as was shown below.
}
The angular components vanish due to the spherical symmetry as is also the case to the fluid velocity $u^\mu$ normalized as $g_{\mu\nu} u^\mu u^\nu=-1$, where $g_{\mu\nu}$ is the metric tensor including the off-diagonal component to be balanced with the radial momentum of matter such that 
\beal{
g_{\mu\nu} \d x^\mu \d x^\nu
= - f (\d x^0)^2 + h (\d r +\psi \d x^0)^2 +r^2 \tilde g_{ij} \d x^i \d x^j, 
\label{LE}
}
where $\psi$ is a newly introduced variable for the radial movement of the reference frame with velocity $\beta=-\sqrt{h/f}\psi$. 

In order to decode the Einstein field equation, $G^\mu\!_\nu=\kappa^2 T^\mu\!_\nu$, in terms of local thermodynamics, the thermodynamic observables are required to be determined suitably. 
To determine a geometric class of thermodynamic observables such as the local temperature $T$ and the thermodynamic local volume element $v$, it is helpful to change the original reference frame of coordinates to an instantaneously rest one by $\check x^0=x^0, \check r=r+\psi x^0$, so that the line element becomes $- f (\d\check x^0)^2 + h \d\check r ^2 +\check r^2 \tilde g_{ij} \d x^i \d x^j$ at $x^0=0$. At this moment, the line element is of the same form as in the previous hydrostatic case, which suggests that the temperature $\check T$ and the volume element $\check v$ observed in this frame can be determined by $\check T=T_\circ/\sqrt f$, $\check v= \sqrt{h\check r^{2(d-2)}\tilde g}$ as in the previous hydrostatic case. By performing the inverse transformation back to the original frame, these quantities receive the special relativistic effect pointwise at $x^0=0$. 
The transformation rule of the temperature specified in  \cite{Ott1963LorentzTransformationDW,Arzelis1965TransformationRD,møller1967relativistic} fixes $T=\check T/\sqrt{1-\beta^2}$. Similarly $v=\check v/\sqrt{1-\beta^2}$. 
Note that the Tolman relation holds exactly in a form $T=T_\circ e^{-\phi}$, where the gravitational potential is currently given by $\phi=\half\log(f-h\psi^2)$.  
On the other hand, a non-geometric class of macroscopic variables are determined from the energy-momentum tensor and the additional current. 
Due to the presence of steady flow, the system becomes anisotropic, so that the pressure perpendicular to the flow direction is $T^i\!_i = \check p$, while that in the radial direction is $p=T^r\!_r = \check p + \wp$, where $\wp = (u^r)^2(-\check\rho-\check p+2u\cd q)/(h^{-1}-f^{-1}\psi^2+2(u^r)^2)$, where the dot means the contraction of two vectors as $u\cd q = u^\mu g_{\mu\nu}q^\nu$. 
The energy density is $\rho=-T^0\!_0 = \check\rho + 2u\cd q -\wp$, while the number density is $j^0=q^0/\chi$. 
Multiplied by the local volume element, the internal energy density is determined as $u=\rho v$ and the thermodynamic number density is $n=j^0v$ in the proper coordinates. Employing these quantities, the entropy density was determined to satisfy both the local Euler's relation $Ts = u + pv -\mu n$ and the first law of thermodynamics $T\d s=\d u + p\d v -\mu \d n$, where $\mu$ is the chemical potential determined to satisfy the equation derived from the Gibbs-Duhem relation $nT\d(\mu/T)=v\d p - (u+vp) \d\log T$ as 
$\mu= -(d-2) T \int \frac{\wp}{q^0}\frac{\chi}T\frac{\d r}r$. 
As seen from this expression, the ratio of the temperature and the chemical potential is not constant any more differently from the hydrostatic case \cite{RevModPhys.21.531}. 
Note that these local thermodynamic relations lead to an equation $p' = ( \rho + p ) T'/T - (d-2)\wp/r$ derived from the covariant conservation equation of the energy-momentum tensor $\nabla_\mu T^\mu\!_r=0$.

Once the thermodynamic observables are properly determined, the Einstein equation can be decoded into the structure equations for the equilibrium system as follows. 
To the end, it is conventional to introduce the so-called gravitational mass or energy within the radius $r$, which is currently defined as $M = (1 - h^{-1}+f^{-1}\psi^2)(d-2) \tilde V r^{d-3}/(2\kappa^2)$, where $\tilde V=\int \d ^{d-2}x \sqrt{\tilde g}$. Then the gradient of the quantity $M$ is computed as 
\be 
\frac{\d M}{\d r} = r^{d-2}\tilde V\rho. 
\ee
On the other hand, the temperature gradient is 
\be 
\frac{\d T}{\d r} =-\frac{\kappa^2}{\tilde Vr^{d-2}} \frac{(d - 3) M + \tilde Vr^{d-1}p}{ d - 2 - \frac{2\kappa^2 M}{\tilde Vr^{d-3}} }T,
\ee
while the pressure gradient is 
\beal{
\frac{\d p}{\d r} =-(\rho + p ) \frac{\kappa^2}{\tilde Vr^{d-2}} 
\frac{ (d - 3) M + \tilde Vr^{d-1}p} { d - 2 -\frac{2\kappa^2 M}{\tilde Vr^{d-3}} } - \frac{d  - 2}r \wp. 
\label{pGradient}
}
The last is the extension of the TOV equation including the effect of steady flow, which is corrected by the difference of the anisotropic pressures of the system denoted by $\wp=p-\check p$ constrained by an additional equation of state in the steady system. 
Note that this form of the contribution of the anisotropic pressure itself was already obtained in earlier research to study the effect of anisotropic pressure to a hydrostatic equilibrium system \cite{1974ApJ...188..657B,1981JMP....22..118C,Bayin:1982vw}. (See a review \cite{Herrera:1997plx} for references therein.)
However, in the earlier research the physical origin of anisotropic pressure in a spherically symmetric hydrostatic equilibrium system was not clarified,%
\footnote{ 
In a case of galactic systems such as a Kapteyn universe, anisotropic pressure was observed and its origin was identified with the anisotropic variance of velocity distribution \cite{1922MNRAS..82..122J}, though such a galactic system has spherical symmetry more or less broken. 
}
while in the recent research its origin is clear due to the nonzero flow momentum of the fluid system.

\subsection{Perturbation around the hydrostatic limit} 
\label{Perturbation} 

Although the above results were obtained exactly without using any approximation, it is instructive and also convenient for later analysis to perform the perturbative analysis of the equilibrium system with steady flow around the hydrostatic limit. 
The small expansion parameters are the radial component of the flow velocity $u^r$ as well as the variable $\psi$ in the off-diagonal component of the metric. Their relation is implicit in a general situation without using any equation of state, while these appear on an equal footing and are counted as the same order in the perturbation. 
For example, the normalization condition imposes the time component of the flow velocity as $u^0=(u_0 - g_{0r}u^r)/g_{00}$ with $u_0 = -\sqrt{- g_{00}(1+ (u^r)^2/g^{rr}) }$, so it is expanded within the next-to-leading order as $u^0 = \frac1{\sqrt{f}}(1+\frac h2 (u^r +\frac\psi{\sqrt{f}} )^2 ) + \cdots$, where the ellipsis represents higher order terms, as asserted.

This can be understood more generally by studying the response for the scale transformation in the radial direction, $r \mapsto \lambda r$ with $\lambda$ positive.
Under the transformation the metric components transform as $\psi \mapsto \lambda\psi$, $h\mapsto \lambda^{-2} h$, $f\mapsto f$, while the flow velocity $u^r \mapsto \lambda u^r$, $u^0 \mapsto u^0$. This suggests that $\psi$ and $u^r$ are expanded in odd power of the small parameters while $f, h, u^0$ are in even power. 
Similarly, $\check\rho, \check p, q^0$ are in even power, while $q^r$ is in odd. Note that the expansion of $q^0$ starts with the quadratic order since the contribution of the current for steady flow vanishes in the hydrostatic limit. 
Accordingly the subleading order of the Einstein equation is quadratic for the diagonal components but is linear for the off-diagonal components. 
In this situation with broken general covariance, the approach of the covariant perturbation, which has been used typically in the analysis of fluid system such as stability \cite{HISCOCK1983466,PhysRevD.31.725}, is not useful any more. 

The linear analysis of the off-diagonal components is done as follows. 
Since $G^r\!_0$ vanishes, the Einstein equation for this component yields a nontrivial constraint such that $q^r = -(\check \rho +\check p +q_0/u_0)u^r= -(\mathring\rho +\mathring p)u^r+\cdots$. 
On the other component, the Einstein tensor is $G^0\!_r = (d -2) \psi  (\log\mathring h+\log\mathring f)'/(2 r\mathring f)+\cdots$, while the energy-momentum tensor $T^0\!_r = (\check\rho+\check p)u^0 u_r+ q_ru^0+\cdots=(\mathring \rho+\mathring p)\mathring h\psi(\mathring u^0)^2+\cdots$, where the above expression of $q^r$ was used. Since $\psi$ is nonzero, the Einstein equation for this component in the leading order is $(d -2) (\log\mathring h+\log\mathring f)'/(2 r\mathring f) =\kappa^2 (\mathring \rho+\mathring p)\mathring h(\mathring u^0)^2$, which is trivially satisfied by using the zeroth order of the Einstein equation obtained from a diagonal component since $(\mathring u^0)^2 =1/\mathring f$. 

For the diagonal components, as mentioned, the linear analysis is not sufficient to derive the small deviation from the hydrostatic limit. Therefore, since we already know the exact results of the structure equations reviewed above, it is direct to derive the correction of each structure variable from them.
The calculation is straightforward and the final result is as follows. 
\beal{
\frac{\d \delta M}{\d r} =& r^{d-2}\tilde V\delta\rho, \\
\frac{\d }{\d r}(\frac{\delta T}{\mathring T}) =&
- \frac{\kappa^2}{\tilde Vr^{d-2}} \bigg\{\frac{(d - 3)\delta M + \tilde Vr^{d-1}\delta p}{ d - 2 - \frac{2\kappa^2 \mathring M}{\tilde Vr^{d-3}} } + \frac{2\kappa^2}{\tilde Vr^{d-3}}  \frac{ (d - 3)\mathring M + \tilde Vr^{d-1}\mathring p }{ ( d - 2 - \frac{2\kappa^2 \mathring M}{\tilde Vr^{d-3}} )^2 } \delta M \bigg\},\\
\frac{\d \delta p}{\d r} =&- \frac{\kappa^2}{\tilde Vr^{d-2}}  \bigg\{ 
\frac{ (d - 3)\mathring M + \tilde Vr^{d-1}\mathring p} { d - 2 -\frac{2\kappa^2\mathring M}{\tilde Vr^{d-3}} }(\delta\rho +\delta p )+(\mathring\rho +\mathring p ) 
\frac{ (d - 3)\delta M + \tilde Vr^{d-1}\delta p} { d - 2 -\frac{2\kappa^2\mathring M}{\tilde Vr^{d-3}} } \nn
&\qquad\quad + (\mathring\rho +\mathring p )\frac{2\kappa^2}{\tilde Vr^{d-3}} 
\frac{ (d - 3)\mathring M + \tilde Vr^{d-1}\mathring p} {( d - 2 -\frac{2\kappa^2\mathring M}{\tilde Vr^{d-3}})^2 } \delta M\bigg\} - \frac{d  - 2}r\grave \wp. 
}
Here $\grave \wp$ is the leading term of $\wp=p-\check p$ and computed as $\grave \wp = -\mathring h(\mathring\rho+ \mathring p) (u^r)^2=-(d - 2)\tilde Vr^{d-3}(\mathring\rho+ \mathring p)  (u^r)^2 /( 1 - 2\kappa^2\mathring M) $. 
Each symbol with $\delta$ prepended describes the deviation from the hydrostatic limit. $\delta M= M - \mathring M, ~\delta T= T - \mathring T$, while the correction of the energy density and the pressure needs some computation and the results are given as follows. 
\beal{
\delta\rho=&\delta\check\rho -2\frac{T_\circ}{T_\vartriangle} j^0+ (1+2\frac{\mathring T}{T_\circ} \frac \psi{u^r} ) \grave\wp, \quad 
\delta p=\delta\check p+\grave\wp,
}
where $\delta\check\rho=\check \rho-\mathring\rho,~ \delta\check p=\check p -\mathring p$ and it was used that $\chi=\mathring T/T_\vartriangle+\cdots$ with $T_\vartriangle$ a positive constant, and that $
u\cd q
= -\frac{T_\circ}{T_\vartriangle} j^0+ (1+\frac{\mathring T}{T_\circ} \frac \psi{u^r} )\grave\wp +\cdots$. The expression of $\chi$ is derived in the next section. 
Interestingly, turning on steady flow gives negative feedback for both density and pressure. 
As seen from the final result, it will be quite hard to be obtained simply by perturbing the Einstein equation from the hydrostatic solution. 

\section{Entropy current} 
\label{EC} 

\subsection{Method} 

Now we move on to determining the entropy current for the self-gravitating equilibrium system with steady flow extending the previous result in the hydrostatic equilibrium system \cite{Yokoyama:2023nld}. 
Previously the entropy current was constructed by refining the method to construct a general conserved current proposed in \cite{Aoki:2020nzm}. 
The method is that if the system has the energy-momentum tensor covariantly conserved, then a conserved current $s^\mu$ can be obtained as $s^\mu=\sqrt{|g|}T^\mu\!_\nu\xi^\nu$ for any vector field $\xi^\nu$ to satisfy a differential equation $T^\mu\!_\nu\nabla_\mu\xi^\nu=0$.
Any Killing vector field is contained in such a class of vector fields and the associated conserved current with a Killing vector leads to a Noether charge. It was argued that there exists a nontrivial vector field different from the Killing ones and the resulting non-Noether conserved charge describes the entropy of the system \cite{Aoki:2020nzm}. 
Such a nontrivial vector field was indeed constructed for the hydrostatic equilibrium system as follows \cite{Yokoyama:2023nld}. 

The proposed prescription in \cite{Yokoyama:2023nld} is to solve the differential equation by setting the ansatz for such a nontrivial vector field to be made of a linear combination of all vector fields consisting of the fluid system.
This ansatz is natural in the regard that the entropy of the system is expected to encode the whole information of macroscopic dynamics.%
\footnote{ 
It would be an interesting problem to prove this ansatz in terms of a microscopic viewpoint, for example, by means of the approach developed in \cite{PhysRevD.99.125011,BECATTINI2024138533}. 
}
As reviewed in section~\ref{ExactResult}, the hydrostatic equilibrium system consists of only one velocity field of a perfect fluid $\mathring u^\mu$. In this situation the ansatz is $\mathring\xi^\mu = -\mathring\zeta\mathring u^\mu$,  where $\mathring\zeta$ is an unknown function to be determined. Then the differential equation boils down to $\mathring\zeta'  = \frac{ \mathring p}{\mathring\rho} \frac{-\mathring\rho' }{\mathring\rho + \mathring p}  \mathring\zeta $, which can be solved as $\mathring\zeta=\frac{T_\circ}{\mathring T}\frac{\mathring\rho+\mathring p}{\mathring p}$ with $T_\circ$ an integration constant. 
Substituting this into the expression of the conserved current leads to $\mathring s^\mu=\mathring v\mathring f(\mathring\rho+\mathring p)\mathring u^\mu/T_\circ$, where $\mathring v=\sqrt{\mathring hr^{2(d-2)}\tilde g}$. As a result the entropy density was obtained by the time component of the entropy current as $\mathring s= (\mathring u +\mathring p\mathring v)/\mathring T$, where $\mathring u=\mathring\rho \mathring v$ is the thermodynamic internal energy density. This is nothing but the Euler's relation holding locally. It was also shown that the entropy density determined in this way satisfies the first law of thermodynamics $\mathring T\d\mathring s=\d\mathring u + \mathring p\d\mathring v$. Note that this result was shown without using any approximation and holds exactly. 

In order to extend this result to the current steady system, recall its energy-momentum tensor given by \eqref{EMT}, which is constituted by not only a single fluid velocity $u^\mu$ but also another current $j^\mu$ accounting for the steady flow. 
Then, following the prescription, we set the ansatz such that $\xi^\mu = -\zeta u^\mu - \varsigma j^\mu$, where $\zeta, \varsigma$ are unknown functions to be determined. 
Now we encounter a problem. Even though we substitute this into the differential equation and solve it as previously, only one of two unknown functions is determined. 
What is another determining condition for the additional unknown function? 
The answer of the author is to {\it reverse the derivation process in the hydrostatic case}. 
That is, another condition to fix the entropy current is to equal its time component to the entropy density determined by satisfying the thermodynamic relations. 
Note that the desired entropy density was already determined in \cite{Yokoyama:2025gbi} as reviewed in section~\ref{ExactResult}, and that this condition clearly holds at the leading order as was shown in \cite{Yokoyama:2023nld}.
For convenience, this condition is referred to as the matching condition.

This matching condition leads to an unconventional form of entropy current as follows. To see it, substituting the ansatz of $\xi^\mu$ into the expression of the entropy current, one finds $s^\mu = a u^\mu + b j^\mu$, where $a= \sqrt{|g|} ( ( \check\rho  - u\cd j ) \zeta +(- u\cd j(\check p+\check\rho) + q\cd j )\varsigma  )$, $b=\sqrt{|g|} ( \zeta+( -\check p + u\cd j) \varsigma)$.
The matching condition imposes the unknown coefficient $a$ to be $a = (s-b j^0)/u^0$, and leads to 
\be 
s^\mu 
= \frac{s-b j^0}{u^0} u^\mu + b j^\mu, 
\label{ExactFormEC}
\ee
as was asserted.%
\footnote{ 
For example, in a standard textbook of relativistic fluid mechanics by Landau-Lifshitz \cite{landau1987fluid}, the entropy flux is given by $\sigma_{\rm (LL)} u^\mu$ for a simple fluid system, in which $\sigma_{\rm (LL)}$ is the entropy per unit proper volume. 
Its invalidity in a general situation may be presumed by recasting the problem in electromagnetism.  
The analogous relation is between electric current $j^\mu$ and charge density $\rho$ such that $j^\mu=\rho v^\mu$. This relation does not hold in general as confirmed in textbooks on electromagnetism. 
}
As a result in order to determine the entropy current it is sufficient to fix the unknown parametric function $b$, and this can be done for $s^\mu$ to satisfy the conservation equation. 
This clarifies the procedure to determine the entropy current in the current steady system exactly. Its implementation is done perturbatively below.

\subsection{Perturbative construction}
\label{PerturbativeConstruction}

Before perturbatively implementing the above method to determine the entropy current, it is convenient to do some preparatory computation.

The first computation is the expansion of the fluid $\vartheta = \nabla\cd u$.
The leading term is already determined as $\lim_{u^r\to0}(\frac\vartheta{u^r}) =\frac{ -\mathring\rho'}{\mathring\rho+\mathring p}$ by the infinitesimal analysis of the relativistic fluid equation around the hydrostatic limit \cite{Yokoyama:2023nld}.
The exact expression in the current system can be obtained from the fluid equation, which is derived from the covariant conservation equation of the energy-momentum tensor, $\nabla_\mu T^\mu\!_\nu=0$. Multiplying the velocity field $u^\nu$ for both sides and contracting the index $\nu$, one obtains
$
\frac{\vartheta}{u^r } =\frac{-\check\rho' + \varepsilon}{\check p+\check\rho},  
$
where $\varepsilon=(\vartheta u\cd q + u\cd \dot\nabla q -\nabla\cd q)/u^r$ with $\dot\nabla:=u\cd\nabla$.

The second is the scalar factor $\chi$ connecting two vectors as $q^\mu=\chi j^\mu$. Its expression can be obtained by combining an equation of state imposed on $j^\mu$. Here the covariant conservation equation is imposed. 
Since the divergence of the current is computed as 
$\nabla\cd j= \chi^{-2}u^r\{ - \iota \chi'  + \chi (\iota' + \iota \vartheta/u^r) \} 
$, where $q^r = \iota u^r$ with $\iota=-(\check p +\check \rho+j_0/u_0)$, the conservation equation for the current boils down to a differential equation $ \chi' = \chi ((\log\iota)' + \vartheta/u^r) $. 
At the leading order, this reduces to $\mathring\chi' = \mathring\chi \frac{\mathring p'}{\mathring\rho+\mathring p} $, where $\mathring\chi$ is the leading term of $\chi$.
Since this differential equation is the same as that of the leading temperature, $\mathring T' = \mathring T \frac{\mathring p'}{\mathring\rho+\mathring p} $, 
it can be solved as $\mathring\chi = \mathring T/T_\vartriangle$, where $T_\vartriangle$ is an integration constant. This integration constant is chosen to be positive since 
$\mathring\chi$ obeys the same differential equation as the temperature at the leading order. 

Now we are ready to determine the parametric function $b$ in the entropy current satisfying the matching condition expressed as \eqref{ExactFormEC} order by order by solving the entropy current conservation equation. 
Note that $b$ is expanded in even power as is the same as the entropy density.

The zeroth order is fixed as follows. As reviewed above in the case of the hydrostatic limit, the entropy current conservation equation was solved at the leading order, and the resulting entropy current was determined as $\mathring s^\mu=\mathring a u^\mu$, where $\mathring a= \sqrt{|\mathring g|}  \mathring\rho \mathring\zeta=\mathring s/\mathring u^0$. 
This result imposes $b$ to vanish at the zeroth order.
Note that since $b$ can be expanded as $b=\sqrt{|\mathring g|} ( \mathring\zeta -\mathring p\varsigma)+\cdots$, the leading term of $\varsigma$ in the ansatz of $\xi^\mu$ is determined as $\varsigma=  \mathring\zeta/\mathring p+\cdots.$

The quadratic order is determined in the following manner. 
To the end, the divergence of the entropy current needs to be computed within the cubic order.
In the current spherically symmetric system, the divergence of the current is computed as $\partial\cd s=\partial_r s^r$, so that it is necessary to compute the radial component of the entropy current within the cubic order. Since $b$ is of the quadratic order, $s^r=\frac s{u^0} u^r +\frac b\chi q^r+\cdots=(\frac {\rho+p-\mu j^0}{u^0} +\frac{T_\vartriangle} vb \iota )\frac vTu^r+\cdots$,  where $\chi=\mathring\chi+\cdots=T/T_\vartriangle+\cdots$. 
Here, any variable used in the unperturbed system is rewritten in terms of its corresponding one in the system with perturbation turned on by adding terms of higher order suitably. 
This technique will be useful to reduce the complexity of the perturbative computation to some extent as seen below.
The caveat to use this technique is that the expression at each order is unique only up to higher order terms, and that it is important not to neglect any term within the order focused in the perturbative calculation at each order once the expression at each order is fixed. 
Taking this technique into account, one can expand the radial component of the entropy current as 
$s^r= s_{(0)}^r  +s_{(1)}^r  + \cdots $, where $s_{(0)}^\mu = \frac{T_\circ}{T}(\check\rho+\check p) \frac vT u^\mu$ and $s_{(1)}^r = \tilde b \iota\frac vT u^r$ with $\tilde b =\frac{T_\vartriangle}{v} b - \frac{T_\circ}T
\frac1{ \check\rho +\check p }\{\frac32 \wp  -\chi u\cd j - (\mu-\frac{T_\circ}{T_\vartriangle}) j^0 \}$.  
Here the expansion of $u^0$ was used in such a form that $u^0 = \frac{T}{T_\circ}\{ 1- \frac{1}{\check\rho+\check p} (\chi u\cd j  + \frac{T_\circ}{T_\vartriangle} j^0 -\half\wp) \} +\cdots$. 
One will see that $s^\mu_{(0)}$, which is written only in terms of the variables in the perturbed system, agrees with $\mathring s^\mu$ in the hydrostatic limit.
Then the divergence of $s_{(0)}^\mu$ is computed as 
$\partial\cd s_{(0)} = \frac{T_\circ }{T}\{ -\wp'-\frac{ (d - 2) }{r} \wp +2(\wp - \chi u\cd j)\frac {T'}T + \varepsilon \} \frac vT u^r +\cdots$, while that of $s_{(1)}^\mu$ is
$\partial\cd s_{(1)} 
= \{ (\tilde b\iota)'+\tilde b\iota\frac \vartheta{u^r} \} \frac vT u^r 
=\{ \tilde b'\iota - \tilde bp' \} \frac vT u^r +\cdots
=( \tilde b\frac T{T_\circ} )' \iota\frac{T_\circ} T \frac vT u^r +\cdots$, where  $\check p' = ( \check\rho +\check p +2\wp -2\chi u\cd j ) T'/T -\wp'- (d-2)\wp/r$ was used.
Summing these up, the divergence of the entropy current is obtained as 
$
\partial\cd s 
=\frac{v}{T} u^r \frac{T_\circ}{T} \{ -\wp'-\frac{ (d-2) }{r} \wp +2(\wp - \chi u\cd j)(\log T) ' + \varepsilon +\iota ( \tilde b\frac T{T_\circ})'\} +\cdots.
$
In order for this to vanish within the cubic order, the expression inside the bracket has to vanish. 
This gives a differential equation, which can be easily solved as  $\tilde b = \frac{T_\circ}{T} \int dr  \frac1{\check\rho+\check p}\{ -\wp'-\frac{ (d-2) }{r} \wp +2(\wp - \chi u\cd j)(\log T) ' + \varepsilon\}$.
Finally, $b$ is determined as 
$b=\frac{T_\circ}{T_\vartriangle} \frac vT [ \frac{1}{ \check\rho +\check p }\{ \frac32 \wp  -\frac T{T_\vartriangle} u\cd j - (\mu-\frac{T_\circ}{T_\vartriangle}) j^0\} + \int dr  \frac1{\check\rho+\check p}\{ -\wp'-\frac{ (d-2) }{r} \wp +2(\wp - \frac T{T_\vartriangle} u\cd j)(\log T) ' + \varepsilon\} ] +\cdots$. 
Remark that the expression of $b$ at the leading order contains higher order terms, though the terms of purely quadratic order are obtained easily.

\section{Discussion} 
\label{Discussion} 

The perturbative analysis of a spherically symmetric self-gravitating equilibrium system with steady flow around the hydrostatic limit has been performed. 
Although its main structure equations were exactly obtained before, the perturbative analysis has been useful to deepen the understanding of rich structure of the system. A technical aspect is that due to the violation of the general covariance in this kind of self-gravitating equilibrium system the standard covariant perturbation approach is not useful anymore. 
In particular, the diagonal components of tensors in the system are expanded in even power of small perturbation parameters while the off-diagonal ones are in odd, which have been necessary to be investigated separately.
The differential equations to determine the correction to the structure variables have been derived, which look complicated enough to see why such an important extension of the structure equations including the TOV equation to the inclusion of steady flow has not been completed even perturbatively.

Subsequently the extension of the entropy current to the system with steady flow has been investigated. The current accounting for steady flow contributes to the entropy current, and due to this contribution, there increases an unknown parametric function in the ansatz of entropy current. 
It has been an problem what is the condition for this unknown parametric function to be determined. 
The answer given in this paper has been to request the agreement of the entropy density determined by satisfying the thermodynamic relations and the charge density read from the entropy current. This matching condition has imposed the entropy current to be of a certain unconventional form with one parametric function that is to be determined by the current conservation equation. 
Taking into account the perturbative structure of the entropy current inherited from that of the tensors, the current conservation equation has been solved perturbatively and the entropy current has been determined within the next-to-leading correction. 
There will be no obstruction to carrying out the perturbative construction for higher order apart from complexity.

As a result, it concludes that the unconventional relation of the entropy current determined in the paper and the entropy density is unavoidable. 
While this conclusion might be something unwelcome to cause confusion, it might be a hint to resolve remaining issues in the field of relativistic fluid mechanics such as the instability of dissipative relativistic fluid mechanics \cite{HISCOCK1983466,PhysRevD.31.725}. See a review \cite{Andersson_2007} for more detail.%
\footnote{ 
The stability analysis involving gravitational interaction often becomes intricate.  
See \cite{Herrera_2020} for a recent result.  
}
As far as the author confirms, the matching condition proposed in the paper has never been imposed in the earlier study of relativistic fluid dynamics.  
It would be interesting to reexamine any earlier result, for example those obtained by requiring the second law of thermodynamics, by imposing the matching condition for the entropy current.

For such a future application, let us confirm in advance for the proposed matching condition to be valid in a general non-equilibrium or dissipative process in curved spacetime beyond standard relativistic field theory.\footnote{ 
This part of the argument is added to answer a question about the scope of the application of the proposal in a peer-review process.  
The author would like to thank an anonymous referee for asking it.  
} 
For the purpose let us fix an arbitrary relativistic system and consider its physical process starting from a local equilibrium state at a time $t=t_1$ and ending in another one at a time $t=t_2$.%
Then the second law of thermodynamics states that the local entropy for a subregion $\Sigma$ in local equilibrium both at $t=t_1$ and $t=t_2$ does not decrease under any such process:\footnote{ 
The existence of such a subregion can be shown as follows. 
Take an arbitrary point $x$ in the given relativistic system. The assumption for the system to be a local equilibrium state at $t=t_1$ ensures the existence of a subregion $\Sigma_1$ around $x$ in local equilibrium. Likewise there exists a subregion $\Sigma_2$ around $x$ in local equilibrium at $t=t_2$. Then if $\Sigma$ is taken to be a subregion included in the intersection of $\Sigma_1$ and $\Sigma_2$, then $\Sigma$ is a subregion in local equilibrium both at $t=t_1$ and $t=t_2$. 
}
\be 
S_\Sigma(t_2) - S_\Sigma(t_1) \geq 0 , 
\label{2ndLaw}
\ee
where $S_\Sigma(t)$ is the local entropy for the subregion $\Sigma$ in local equilibrium at the time $t$. 

Adopting the definition proposed in \cite{Yokoyama:2023nld}, the local entropy for the subregion $\Sigma$ at a time $x^0=t$ for a general relativistic field theory on curved space is obtained as 
\be 
S_\Sigma(t) = \int_{\Sigma} {\mathrm d}^{d-1}\vec x \, s(t, \vec x), 
\label{Entropy}
\ee
where $s$ is the entropy density determined as the time component of the entropy current from the matching condition: 
\be 
s(t, \vec x) = s^0(t, \vec x). 
\label{Matching}
\ee
Then the lefthand side of \eqref{2ndLaw} is rewritten as 
\beal{
S_\Sigma(t_2) - S_\Sigma(t_1)
=& \int_{t_1}^{t_2} dt \frac d{dt} S_\Sigma(t)
\nn
=& \int_{t_1}^{t_2} dx^0 \int_{\Sigma} {\mathrm d}^{d-1} x \, \partial_0 s^0(x^0, \vec x) \nn
=& \int_{t_1}^{t_2} dx^0 \int_{\Sigma} {\mathrm d}^{d-1} x \, \partial_\mu s^\mu (x)
\label{Difference}
}
where in the first equality the fact was used that the entropy is a state quantity and in the last that there is no boundary contribution of the entropy current as is usually assumed in relativistic field theory.  
The second law of thermodynamcis \eqref{2ndLaw} states that the quantity \eqref{Difference} is non-negative for any dynamical process with $t_2 \geq t_1$ and any subregion $\Sigma$ in local equilibrium. This implies that the integrand is non-negative for an arbitrary point $x$: 
\be 
\partial_\mu s^\mu (x) \geq 0. 
\ee
As a result it concludes that the matching condition is consistent with the second law of thermodynamics and is applicable to a general process such as a non-equilibrium and a dissipative one in curved spacetime. 

This research hopefully contributes to new progress in the future research of relativistic astrophysics, thermodynamics and fluid dynamics. 

\bibliography{EC}

\end{document}